\documentclass[11pt]{iopart}
\usepackage{amsfonts}
\usepackage{amssymb}
\usepackage{graphicx}
\usepackage{dcolumn}
\usepackage{epstopdf}
\usepackage{bm}
\usepackage{color}

\begin{document}

\title{Non-locality of Two Ultracold Trapped Atoms}

\author{Thom\'{a}s Fogarty and Thomas Busch}
\address{Department of Physics, University College Cork, Cork, Republic
  of Ireland}
\author{John Goold}
\address{Centre for Quantum Technologies, National University of Singapore, 3 Science Drive 2, 117543, Singapore}
\date{\today}
\author{Mauro Paternostro}
\address{School of Mathematics and Physics, Queen's
University Belfast, Belfast, BT7 1NN, United Kingdom}
\date{\today}

\begin{abstract}
  We undertake a detailed analysis of the non-local properties of
  the fundamental problem of two trapped, distinguishable neutral
  atoms which interact with a short range potential characterised by an s-wave scattering length. We show that this interaction
  generates continuous variable (CV) entanglement between the external
  degrees of freedom of the atoms and consider its behaviour as a
  function of both, the distance between the traps and the
 magnitude of the inter-particle scattering length. We first quantify
  the entanglement in the ground state of the system at zero
  temperature and then, adopting a phase-space approach, test the
  violation of the Clauser-Horn-Shimony-Holt inequality at zero
  and non-zero temperature and under the effects of general
  dissipative local environments.
\end{abstract}


%
% 32.80.Pj - Optical cooling of atoms; trapping.
% 03.67.Mn - Entanglement production, characterization, and manipulation.
% 03.65.Ge - Solutions of wave equations: bound states.
% 05.30.Jp - Boson systems.
%

\maketitle

\section{Introduction}

Ultracold atoms have recently emerged as ideal systems for the
exploration of fundamental effects in quantum mechanics, quantum
information and quantum simulation~\cite{Bloch:08}. While a large
amount of attention so far has been directed towards the exploitation
of entanglement in the internal degrees of freedom of the atomic
systems, the exploration of the external degrees of freedom as
valuable physical supports for the encoding of continuous variables
(CV) quantum information has not been considered extensively. This may have been
motivated, arguably, by the difficulties faced so far in achieving a
strong enough interaction between neutral, atomic CVs. However, it is
by today experimentally possible to greatly enhance such coupling by,
for example, driving Feshbach resonances using external magnetic
fields~\cite{Feschbach}. Furthermore, techniques to trap, cool and
control single neutral atoms have improved to an extent that
high-fidelity measurements on single quantum particles are now
possible~\cite{Beugnon:06,Anderlini:07,Wilk:10}. For neutral atoms,
optical lattices and dipole traps have been used in proposals for the
implementation of fundamental two qubit gates
\cite{Jaksch:99,Calarco:00,Isenhower:10}. Moreover, the flexibility
typical of optical potentials allows one to consider spin dependent configurations \cite{Bloch:03}. Another example of this is the ability to shape the trapping geometry in different spatial directions such that the dynamics of the
system along one or several directions can be inhibited. For one-dimensional configurations, interactions can also be significantly
enhanced through so-called confinement-induced resonances~\cite{Olshani:98}.

With all this in mind, we will in the following investigate the
entanglement generated among external degrees of freedom in the
presence of inter-atomic couplings. For this we study an
analytic, one-dimensional model of two atoms in separate harmonic
traps which interact via a pseudopotential \cite{Huang:57}. This model
is an extension of the problem of two atoms interacting in a single
harmonic trap \cite{Busch:98}, and its analytical solution was
recently given in \cite{Krych:09,Goold:10}. For cold atomic systems one-dimensional models
are known to be in good agreement with experimental
results~\cite{Stferle:06}, and the existence of analytical solutions
has made them a good model to be used as a testbed for realistic
studies of various facets of entanglement \cite{Mack:02, Buss:07, You:06,
  Murphy:07, Goold:09}.

Here we extend these works and investigate the non-local nature of the
CV atomic state generated by the interaction between two atoms using a phase-space-based
version of Clauser-Horne-Shimony-Holt (CHSH) inequality derived in
Ref.~\cite{Banaszek:99}. The analyticity of our model allows us to
extend the study to finite temperature and take losses into account, thus providing a full-comprehensive theoretical
characterization of non-classical correlations and
paving the way to their experimental demonstration.

The presentation of the work is organised as follows. In
Sec.~\ref{sec:model_hamiltonian} we introduce the system, briefly
review its exact solution and quantify the degree of entanglement for
its ground state using the von-Neumann entropy. In
Sec.~\ref{sec:wigner} we calculate the Wigner function of the atomic
state and show that it has a considerable negative part, which is a
strong indication of inherent non-classicality. In Sec.~\ref{sec:chsh}
we test for this non-locality by calculating a continuous variable
CHSH-like function~\cite{Banaszek:99} and discuss and illustrate the
violation of local realistic theories for a wide range of parameters.
In order to connect the results to experiments we extend our study to
include the effects of a general dissipative environment in
Sec.~\ref{sec:LossyDetector}. Finally, Sec.~\ref{sec:Conclusions}
draws our conclusions and accesses the impact of our work.

\begin{figure}[tb]
\begin{center}
  \includegraphics[width=\linewidth,bb= 0 0 1261 483]{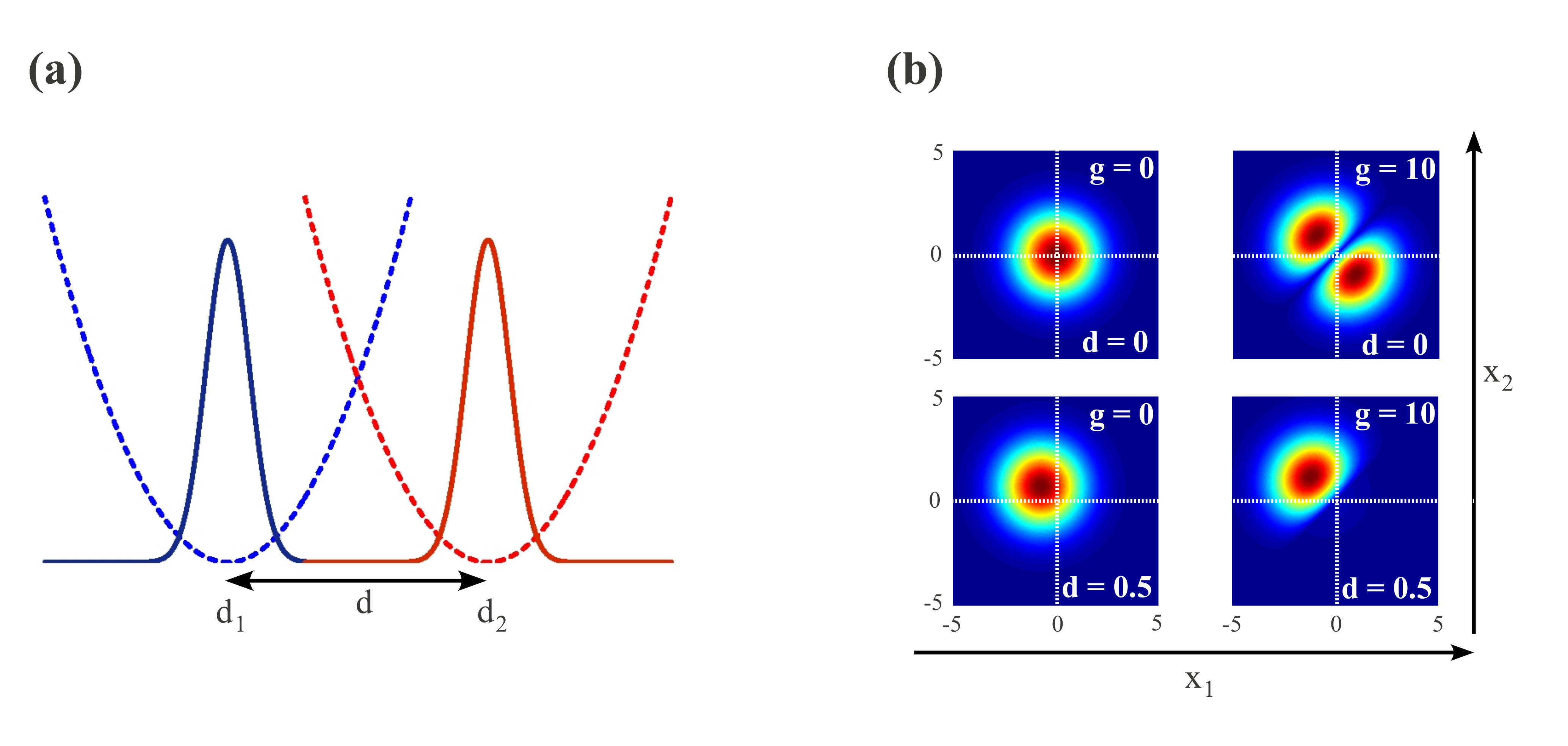}
  \end{center}
  \caption{Panel {\bf (a)} shows a schematic representation of the
    system at hand [see the Hamiltonian model in Eq.~\ref{Hamiltonian}]. Panel {\bf (b)}
    shows the two-particle probability density for two different
    distances $d$ between the traps and two distinct values of the scaled interaction strength $g$. The axes are scaled in terms of $a$ as defined in the text.}
\label{fig:schematic}
\end{figure}

%%%%%%%%%%%%%%% 
\section{Model Hamiltonian}
\label{sec:model_hamiltonian}

The model we consider consists of two bosonic atoms confined along the
$x$ axis (the {\it axial direction}) with two separate, but
overlapping harmonic potentials, as shown in
Fig.~\ref{fig:schematic}{\bf(a)}. The atoms are tightly confined along
directions perpendicular to $x$ ({\it the transverse directions}) by
high-frequency harmonic trapping potentials. As a result of the large energy
level separation associated with the transverse confinement, at low
temperature the transverse motion is restricted to the lowest
mode. The system can then be described by the quasi one-dimensional
Hamiltonian
\begin{eqnarray}
\label{Hamiltonian}
\hat H=&-\frac{\hbar^2}{2m_1}\nabla^2_1-\frac{\hbar^2}{2m_2}\nabla^2_2
    +\frac{m_1}{2}\omega^2(x_1-d_1)^2+\frac{m_2}{2}\omega^2(x_2-d_2)^2\nonumber\\
    &+g_{1D}\delta(x_1-x_2),
\end{eqnarray}
where $m_1$ and $m_2$ are the masses of the two atoms and $x_1$ and
$x_2$ are their respective spatial coordinates. We assume both traps to
have the same frequency $\omega$ and be displaced by the distances
$d_1$ and $d_2$ from the origin of the coordinate system. We model the
atomic interaction using the standard point-like pseudo-potential and
restrict ourselves to s-wave scattering. At low temperatures the
scattering strength is then known to be given by $g_{1D} = -2\hbar^2/m_r
a_{1D}$, where $m_r{=}m_{1}m_{2}/(m_{1}{+}m_{2})$ is the reduced atomic
mass and $a_{1D}$ is the one-dimensional scattering length related to
the actual three-dimensional one via
$a_{1D}{=}-a_{\perp}^{2}/2a_{3D}(1 - Ca_{3D}/a_{\perp})$. Here
$a_\perp$ is the size of the single-atom ground state wavefunction in
the transversal direction and $C\simeq 1.4603\ldots$ is a
constant~\cite{Olshani:98}. By introducing the centre of mass
coordinate $X=(x_1+x_2)/2$ and the relative coordinate
$x=(x_1-x_2)/2$, the two-atom wavefunction can be factorised into
$\phi(X)\psi(x)$ with $\phi(X)$ [$\psi(x)$] being the wavefunction for
the centre-of-mass [relative motion] dynamics. Correspondingly, the
Schr\"odinger equation decouples as
\begin{eqnarray}
   \label{eq:Modelhamiltonian}
  \left(-\frac{\hbar^2}{2M}\frac{\partial^2}{\partial X^2}
        +\frac{1}{2}M\omega^2X^2\right)\phi(X)
        =\hbar\omega\left(n+\frac{1}{2}\right)\phi(X)\;,\\
          \label{eq:ModelhamiltonianRel}
  \left (-\frac{\hbar^2}{2m_r}\frac{\partial^2}{\partial x^2}
         +\frac{1}{2}m_r\omega^2(x-d)^2
         +g_{1D}\delta(x)\right)\psi(x)
         =\hbar\omega\left(\nu+\frac{1}{2}\right)\psi(x)\;,        
 \end{eqnarray}
 where we have taken $m_1{=}m_2{=}m$ for simplicity,and defined $d=d_1-d_2$, $M=2m$ and $m_r=m/2$. Clearly, the centre-of-mass dynamics
 has the form of simple harmonic motion while the relative problem
 consists of a displaced harmonic oscillator subjected to a point-like
 disturbance at the origin of the coordinate system.

\begin{figure}[tb]
\begin{center}

\includegraphics[width=6in]{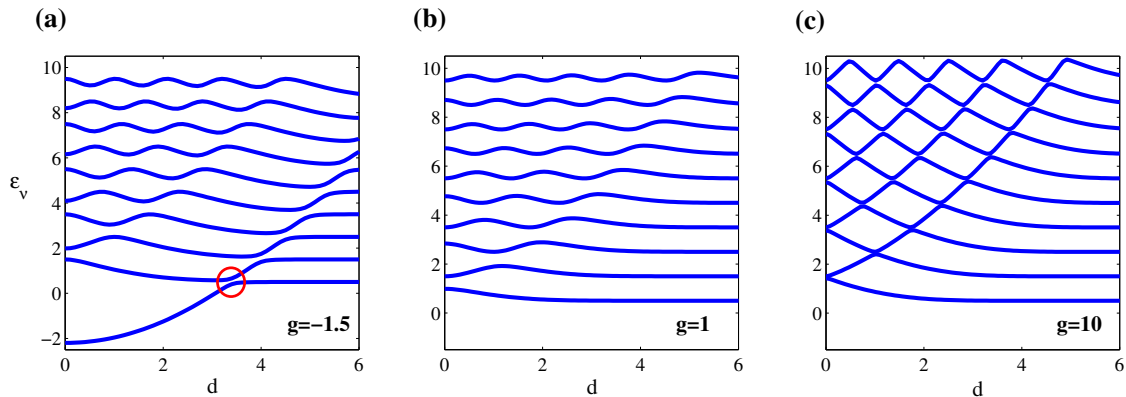}

\end{center}
\caption{Energy versus distance $d$ between the traps for different values of the scaled interaction strength {\bf(a)} $g=-1.5$, {\bf(b)} $g=1$ and {\bf(c)} $g=10$. For large values of $g$ the energy levels become degenerate at $d=0$. For finite distances between the traps resonances appear whenever two levels approach each other. For a repulsive interaction, the ground state is not affected by such resonances, however, for an attractive interaction, a resonance can be observed (red circle in panel {\bf (a)}). }
\label{pic:energyspec}
\end{figure}

 For the sake of completeness, in the following we will briefly sketch the
 steps required to solve Eq.~(\ref{eq:ModelhamiltonianRel}). Our
 approach follows the detailed treatment given in
 Refs.~\cite{Krych:09,Goold:10}. For simplicity of notation we first scale all the 
 lengths in units of $a=\sqrt{\hbar /m \omega}$, which is the width
 of the ground state wavefunction {for a single unperturbed particle of mass $m$}  along the axial direction {of the harmonic trap}, and all energies in units of $\hbar\omega$.
  Eq.~(\ref{eq:ModelhamiltonianRel}) thus becomes (for $x\neq 0$)
\begin{equation}
  \frac{d^2\psi}{d\xi^2}+\left(\nu+\frac{1}{2}-\frac{\xi^2}{4} -g\,\delta(\xi+d) \right)\psi=0\;,
\end{equation}
where $g=g_{1D}a/(\hbar\omega)$ is the renormalised strength of the $\delta$-barrier, $\xi=(x-d)$ is a shifted spatial coordinate and we have dropped the
spatial dependence of $\psi$ for convenience. The solutions to this
equation can be given in terms of parabolic cylinder functions,
$D_{\nu}(\pm\xi)$, of order $\nu$ as follows
\begin{eqnarray}
%\begin{split}
  \psi_l&=N_lD_{\nu}(-\xi)%&=N_lD_{\nu}(d-x)
  &\quad\mbox{for }\;x<0,\\
  \psi_r&=N_rD_{\nu}(\xi)%&=N_rD_{\nu}(x-d),
  &\quad\mbox{for } \;x>0,
%\end{split}
\end{eqnarray}
where $N_l$ and $N_r$ are the normalization factors that can be
calculated by imposing continuity of the solutions at the position of
the $\delta$-function. Explicitly, this leads to the following
conditions
\begin{eqnarray}
\frac{1}{2}\nu\left[D_{\nu-1}(-d)D_{\nu}(d)
                   +D_{\nu-1}(d)D_{\nu}(-d)\right]
      -g D_{\nu}(-d)D_{\nu}(d)=0\;,
\end{eqnarray}
for non-zero value at the position of the $\delta$-function and
\begin{equation}
  \label{eq:sol2}
  N_{r} \nu D_{\nu-1}(-d)+N_{l} \nu D_{\nu-1}(d)=0\;,
\end{equation}
whenever the functions are zero at the $\delta$-function. This also
determines the energy spectrum of the system, which exhibits
trap-induced shape resonances due to energy-level
repulsion, as shown in Fig.~\ref{pic:energyspec}~\cite{Krych:09,Goold:10}. Notably, a resonance in the ground state only appears for $g<0$, which is due to the existence of a bound state in this situation (indicated by circle in Fig.~\ref{pic:energyspec}{\bf(a)}). 
The ground-state wavefunction can
now be obtained as $\Psi_0(x_1,x_2)=\phi(X)\psi(x)$ and on the
right-hand side of Fig.~\ref{fig:schematic} we show its two particle
probability density, $|\Psi_0(x_1,x_2)|^2$. The repulsive interaction between the particles is evident as a zero line along the diagonal in the probability density when $x_{1}=x_{2}$. For a finite trap separation the particles become localised in their respective traps and the two particle probability density moves to occupy the upper left-hand side quadrant.

\begin{figure}[tb]
\begin{center}
\includegraphics[width=6in]{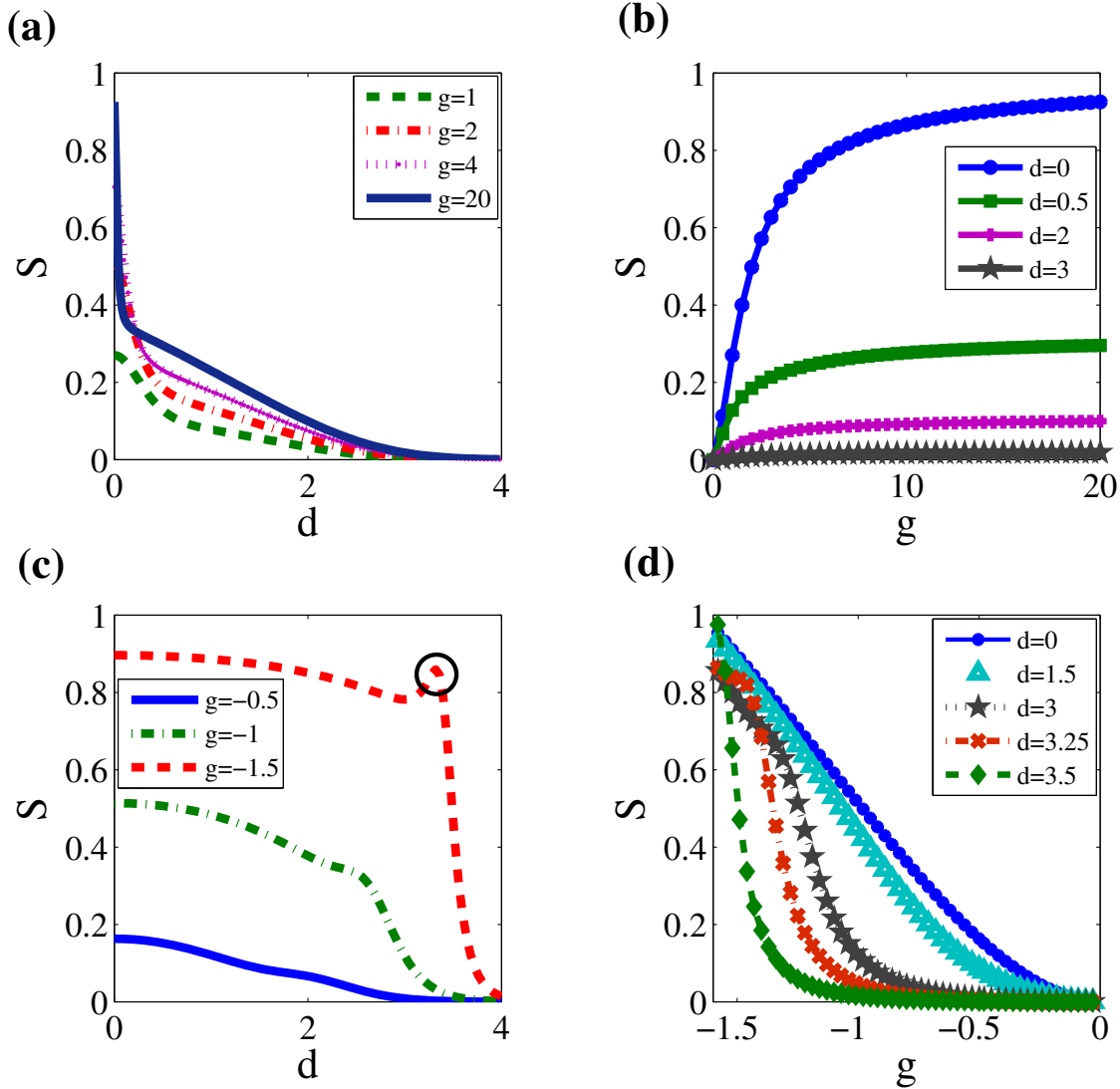}
\end{center}
\caption{Von Neumann entropy of the ground state for repulsive
  [panels {\bf (a)} and {\bf (b)}] and attractive interaction
  [panels {\bf (c)} and {\bf (d)}]. Plots are shown for the von
  Neumann entropy versus trap separation, {\bf (a)} and {\bf (c)}, and
  von Neumann entropy versus particle interaction strength, {\bf (b)}
  and {\bf (d)}. The local maxima visible in {\bf (c)} and {\bf (d)} for
  certain values of $d$ are connected to the appearance of shape
  induced resonances in the energy spectrum, as seen in Fig.~\ref{pic:energyspec}{\bf(a)} for $d \approx 3.5$.}
\label{pic:vonNeumann}
\end{figure}

From this ground state we are able to explore the zero-temperature
quantum correlations of the system by using the von Neumann entropy
$S$ of the reduced single-particle density matrix $\rho_1(x,x')$,
which is determined as the kernel of the reduced density operator
in configuration space
\begin{equation}
  \rho_1(x,x')=\int_{-\infty}^{\infty}\Psi_0(x,x_2)\Psi^*_0(x',x_2)dx_2\;.
\end{equation}
In order to evaluate $S$ we need the eigenvalues $\lambda_i$ of
$\rho_1(x,x')$, which are found by numerically solving the
integral-value equation
\begin{equation}
 \label{eq:rspdm_diag}
 \int_{-\infty}^{+\infty}\rho_1(x,x') \psi_i(x')dx' = \lambda_i\psi_i(x)\;,
\end{equation}
where the $\psi_i(x)$ are the eigenstates associated with the
$\lambda_i$. The von Neumann entropy is then calculated as $S =
-\sum_{i} \lambda_{i} \log_2 \lambda_{i}$ \cite{Mack:02,You:06,Murphy:07}. In Fig.~\ref{pic:vonNeumann} we show $S$ as a
function of both the trap distance $d$ and the interaction strength
$g$. For the case of a repulsive interaction, it can be seen from
Fig.~\ref{pic:vonNeumann}{\bf(a)} that the von Neumann entropy
decreases with increasing {trap} separation. This should be expected
as the short range interaction becomes less important
and the state of the system tends towards the product state of two
non-interacting particles. Fig.~\ref{pic:vonNeumann}{\bf(b)} shows the
behaviour of $S$ as a function of the interaction strength, revealing
that, after an initial raise, $S$ saturates to an asymptotic value
that decreases as $d$ grows. This is again due to the short-range nature of
the interaction potential: as the interaction is ineffective for large
$d$, the steady value of $S$ would be smaller for increasing values of
the separation. For attractive interactions the
situation is slightly different and local maxima and saddle points in
$S$ can be observed at certain values of the trap separation [see
Figs.~\ref{pic:vonNeumann}{\bf(c)} and {\bf(d)}]. A comparison between Figs.~\ref{pic:energyspec} and \ref{pic:vonNeumann}{\bf(c)} shows the correspondence of the appearance of these stationary points and the existence of the above-mentioned trap induced shape resonances for bound states in the energy spectrum. Such simultaneous occurrences have been observed in all our simulations, and based on their strong numerical evidence we conjecture a relation between stationarity in the von Neumann entropy under attractive potentials and trap-induced shape resonances. While such a connection is rather interesting, it goes beyond the scopes of our work and we reserve to investigate it more deeply in the future. Note that, for a given value of $d$, comparatively smaller
values of $g$ are required in the $g<0$ case than in
the repulsive one in order to achieve large values of $S$.

\begin{figure}[tb]
  \includegraphics[width=6in,bb= 0 0 1077 627]{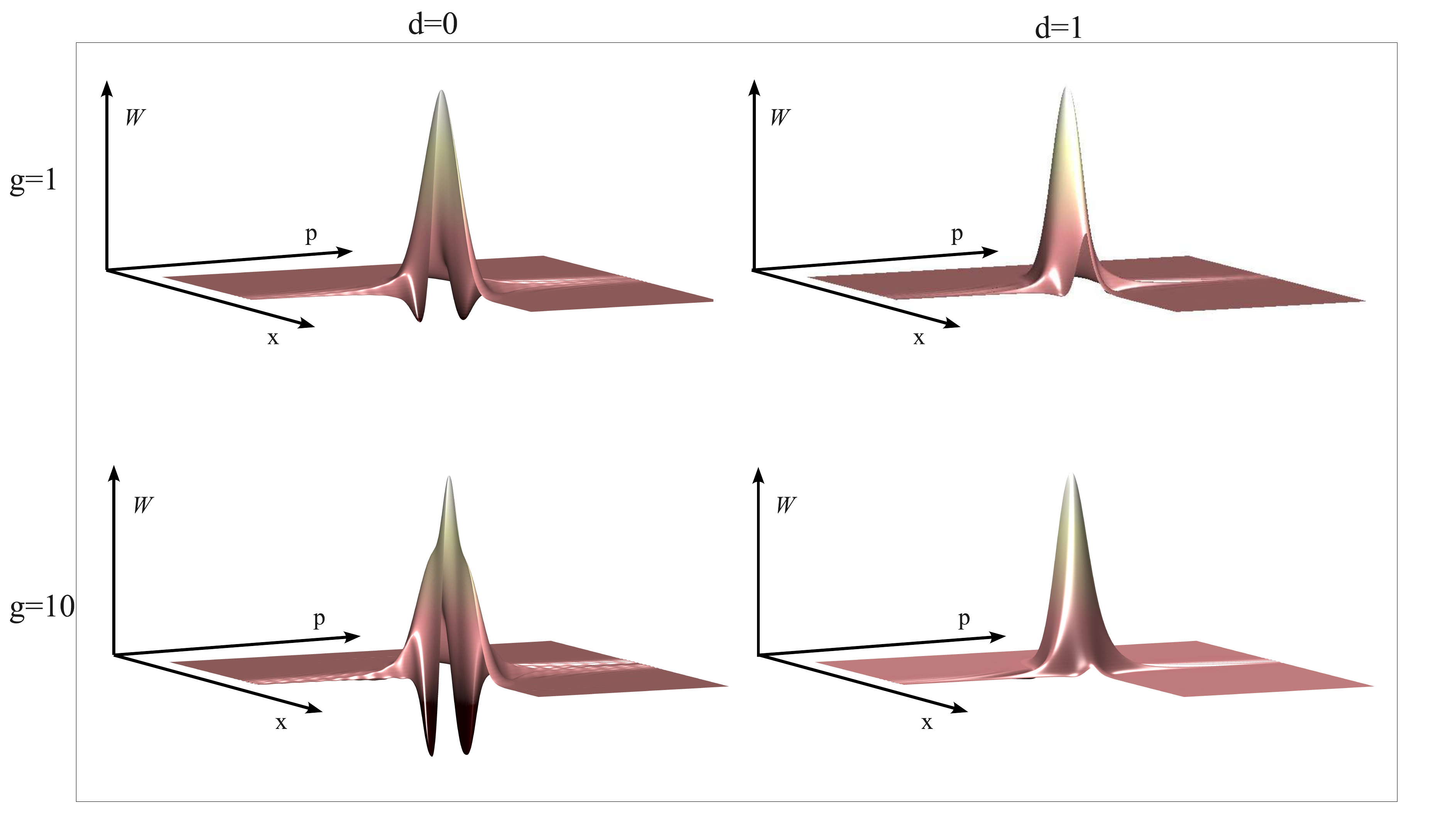}
  \caption{Wigner function for an interaction strength of $g=1$ and
    $g=10$ at trap separations $d=0$ and $d=1$. A quadrant is
    removed from the plot to show the negative parts of the Wigner
    distribution which is symmetric about $p=0$. For $g=1$ a reduction
    in the negative part of the Wigner function is evident for $d=1$
    compared to $d=0$.  For $g=10$ the large negative contribution and
    sharp peak are indicative of the larger interaction strength at
    $d=0$. For $d=1$ the negative volume is significantly less.}
\label{fig:WFg1}
\end{figure}
%%%%%%%%%%%%

\section{Calculation of the Wigner function and assessment of its
  negativity}
\label{sec:wigner}

\begin{figure}[tb]
  \includegraphics[width=6in]{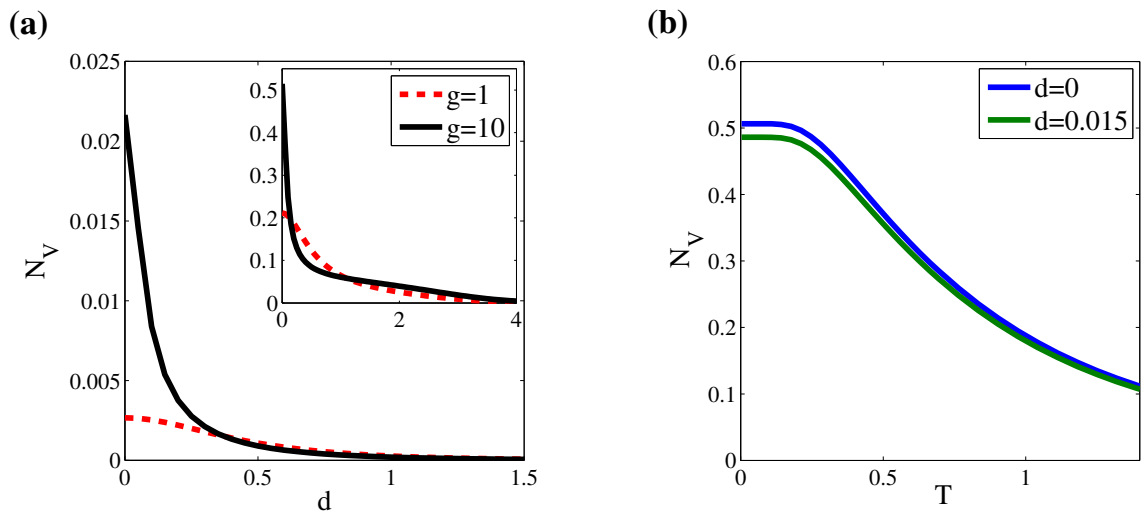}
  \caption{Panel {\bf (a)} shows the negative volume of the complete Wigner
    distribution at zero temperature as a function of $d$. The
    inset shows $N_V$ when only considering the contribution to 
    the Wigner function associated with the relative part of the problem. Panel {\bf (b)}
    shows the negative volume against increasing temperature (measured in units of $\hbar \omega / k_{B}$) for the
    relative part of the  Wigner distribution with an interaction strength of
    g=10.}
\label{pic:negA}
\end{figure}

We will now investigate the non-classicality of the two-atom state in
a much broader range of operative conditions, including finite
temperature. For this, the main tool in our study will be the Wigner
function associated with the two-particle state. For
the specific case at hand, the Wigner function depends on the position
and momentum variables $x_j$ and $p_j$ ($j=1,2$) and is defined
as~\cite{Wigner:32}
\begin{eqnarray}
   W(x_1,p_1;x_2,p_2){=}\!
       \int& d\xi d\varsigma \frac{e^{-\frac{i}{\hbar} p_1\xi-\frac{i}{\hbar}p_2\varsigma}}{4\pi^2\hbar^2}\rho\left(x_1{+}\frac{\xi}{2},x_2{+}\frac{\varsigma}{2},
                      x_1{-}\frac{\xi}{2},x_2{-}\frac{\varsigma}{2}\right).
\end{eqnarray}
By integrating out the momenta or positions one can calculate the
marginal spatial or momentum distributions of the two particles,
respectively. It is straightforward to include the effects of a
non-zero temperature by weighting the higher-order states of the
two-atom spectrum with the appropriate Boltzmann factors,
$P_{n,\sigma}=\frac{1}{{\cal Z}}e^{\frac{-E_{n,\sigma}}{k_BT}}$, where
the $E_{n,\sigma}$ are the energies of the atomic eigenstates
identified by the centre-of-mass and relative-motion quantum numbers
$n$ and $\sigma$, respectively. Moreover, we have introduced the
equilibrium temperature of the system $T$, the Boltzmann constant
$k_B$ and the partition function ${\cal Z}$. We thus get
\begin{equation}                                                                                                                                                                                                                                                                                                                                                                                                                                                                                                                                                                                                                                                                                                                                                                                                                                                                                                                                                                                                                                                                                                                                                                                                                                                                                                                                                                                                                                                                                                                                                                                                                                                                                                                                                                                                                                                                                                                                                                                                                                                                                                                                                                                                                                                                                            
  W(\alpha;\beta)=\displaystyle\sum_n^\infty \displaystyle\sum_{\sigma}^\infty 
                  P_{n,\sigma}W_{n,\sigma}(\alpha;\beta),
\end{equation}

where, for easiness of notation, we have written the Wigner function in terms
of the two quadrature variables $\alpha=(x_1+ip_1)/\sqrt{2}$ and
$\beta=(x_2+ip_2)/\sqrt{2}$. It is widely accepted that the appearance
of negative values in the Wigner function of a system is a strong
indication of non-classicality of the associated state. In fact, in
this case $W(\alpha;\beta)$ cannot be interpreted as a classical
probability distribution describing a microstate in the phase space. Starting from such premises, Kenfack and Zyczkowski have proposed  to use the volume
occupied by the negative regions of $W(\alpha;\beta)$ as a
quantitative indicator for non-classicality (in the sense described above)\cite{Kenfack:04}. Such a (dimensionless) figure of merit 
can be evaluated as 
\begin{equation}
N_V=\frac{1}{2}\left(\int_\Omega|W(\alpha;\beta)|d\Omega-1\right)
\end{equation}
with $\Omega$ being the whole phase-space and $d\Omega=dx_1dx_2dp_1dp_2$. Note that in our case the centre-of-mass part of the wavefunction does not depend on the
interaction between the particles. Therefore, it does not contribute to the
degree of non-classicality and in Fig.~\ref{fig:WFg1} we show the Wigner functions
associated with only the relative part of our problem for two different values of $g$ and $d$.
Negative parts are clearly visible for small values of $d$ and become
more prominent for increasing interaction strength. This is also
visible in Fig.~\ref{pic:negA}{\bf(a)}, where $N_V$ is
plotted against $d$. However, the degree of non-classicality decreases
faster for a larger interaction strength when the traps are moved
apart. The temperature dependence of the negative volume is displayed
in Fig.~\ref{pic:negA}{\bf(b)}, where one can see a very fast decrease
once the system is able to access states beyond the ground state.

\section{Testing non-locality in phase space}
\label{sec:chsh}

The results of the previous Section indicate that a considerable
degree of non-classicality might be set in the state of the external
degrees of freedom of the two trapped atoms, resilient to some extents
to the effects of finite temperature. Moreover, as it should also be
clear from Eq.~(\ref{Hamiltonian}), our study has shown the evident
non-Gaussian nature of the atomic state (as witnessed by the features
of the Wigner function).  While correlations in Gaussian states are
well and easily characterized, we face the lack of necessary and
sufficient criteria for the quantification of entanglement in
non-Gaussian states. In fact, the available entanglement measures for
CV states are based (to the best of our knowledge) on the use of the
negativity of partial transposition criterion formulated in terms of
covariance matrices, which carry exact information on the state of a
system only in the Gaussian scenario~\cite{Braunstein:05}. Interesting 
criteria based on the use of high-order correlation functions of multi-mode bosonic systems have been proposed, recently~\cite{vogel}. However, the experimental determination of such high-order moments can be cumbersome, requiring multi-port interferometric settings~\cite{vogel}. Since here we would like to provide an operatively feasible test for entanglement in the state of the system at hand we will in the following assess non-classicality in terms of non-locality probed in the phase-space of the system studied here. 

We thus consider the CV version of CHSH inequality developed in
Ref.~\cite{Banaszek:99} and will briefly remind the reader of the key
points for completeness.  It is well known that the Wigner function
calculated at the origin of phase space is equivalent to the
expectation value
$W(\alpha=0;\beta=0)=\frac{4}{\pi^2}\langle\hat\Pi_{1}\otimes\hat\Pi_{2}\rangle$,
where $\hat{\Pi}_j$ is the parity operator for mode $j=1,2$
~\cite{Royer:77}.  The total Wigner function can therefore be written
by using {\it displaced parity operators} as~\cite{Royer:77}
\begin{equation}
  W(\alpha;\beta)=\frac{4}{\pi^2}\langle
        \hat{D}_1(\alpha)\hat\Pi_1\hat{D}_1^\dagger(\alpha)\otimes
        \hat{D}_2(\beta)\hat\Pi_2\hat{D}_2^\dagger(\beta)\rangle,
 \label{eq:ExpVal}
\end{equation}
where $\hat{D}_j(\alpha)$ is a displacement operator for mode $j$ of
amplitude $\alpha$~\cite{Braunstein:05}.  A CHSH-like function can
then be built starting from the above as
\begin{equation}
  \label{eqn:chsh}
  \mathcal{B}=\frac{\pi^2}{4}[W(0;0)+W(\sqrt{\mathcal{J}};0)
                      +W(0;-\sqrt{\mathcal{J}})
                      -W(\sqrt{\mathcal{J}};-\sqrt{\mathcal{J}})]
\end{equation}
with ${\cal J}$ a positive constant. Local realistic theories impose
$|{\cal B}|\le{2}$~\cite{Banaszek:99} and any value outside this range
indicates non-local behaviour.

\begin{figure}[tb]
\begin{center}
\includegraphics[width=6in , clip]{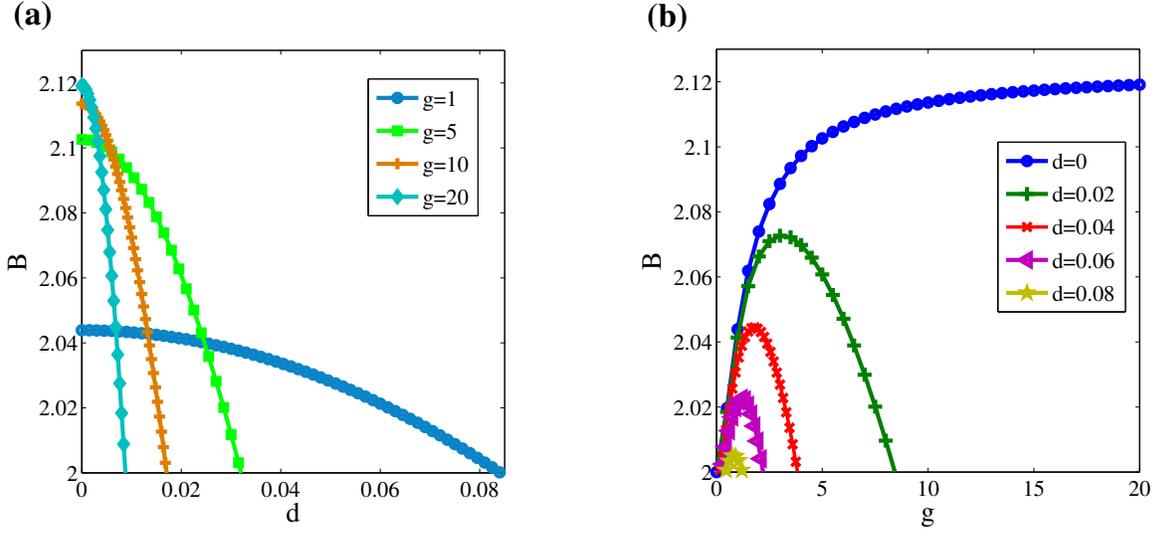} 
\end{center}
\caption{The violation of the CHSH inequality at $T=0$ is shown in
  panel {\bf (a)} against trap separation for various interaction
  strengths and in {\bf (b)} versus $g$, for increasing trap
  separations.}
\label{fig:CHSH0K}
\end{figure}

Equipped with these tools, we can now quantitatively study the
non-locality in the state of our system. Using the Wigner function
calculated in Sec.~\ref{sec:wigner}, we determine the violation of the
CHSH inequality optimised over $\mathcal{J}$ and study the behaviour
of ${\cal B}$ against the interaction strength between the particles
and the distance between the traps. In Fig.~\ref{fig:CHSH0K}{\bf(a)}
we show the numerically optimised values of $\mathcal{B}$ against $d$
for various interaction strengths $g$ and at zero temperature. Clearly,
for short distances the violation of the local realistic bound is
larger for strong interactions. The situation is somehow reverted at
large distances, where weakly interacting atomic pairs appear to
violate the CHSH inequality more significantly. Such an apparently
counterintuitive result can be understood by reminding one that
the one-dimensional interaction strength is inversely proportional to
the one-dimensional scattering length (see
Sec.~\ref{sec:model_hamiltonian}): a lower value of $g$ corresponds to
a larger scattering length. This means that while the correlations steaming from the reduced dimension decay with increasing distance, the influence of the scattering length persists for larger values of $d$. Comparing these results to the von Neumann
entropy shown in Fig.~\ref{pic:vonNeumann} it is evident that
achieving a non-zero von Neumann entropy does not necessarily
correspond to the violation of CHSH inequality, in qualitative agreement with the findings of Ref.~\cite{Mack:02}. It would be interesting to compare the behaviour found
here with those corresponding to longer-range interaction potentials
which might well lead to sustained non-locality at larger
distances. In Fig.~\ref{fig:CHSH0K}{\bf (b)} we show ${\cal B}$ as a
function of the interaction strength. The non monotonic behaviour of
the CHSH function against the interaction strength, as well as the
disappearance of any violation at finite values of $g$ and for
$d\neq{0}$, are striking. It can be understood by realising that the
offset between the traps breaks the symmetry of the system and a large
repulsive interaction between the particles results in less overlap
and therefore less correlations in the phase space. Noticeably, although the CHSH inequality is only violated for $d\lesssim 0.08$, recent experiments involving optical lattices have demonstrated the possibility to off-set atomic trapping potentials with an accuracy of exactly this order of magnitude~\cite{Jessen:09}.

For the case of non-zero temperature we plot the violation of the CHSH
inequality for two values of interaction strengths ($g=1,10$) in
Fig.~\ref{fig:CHSHFT}. While the similarity
of the plots shows the general trends of decay of the correlations
with increasing temperature and distance, one can note that for $g=1$
the system is more resilient to the effects of an increasing
temperature than in the stronger-interaction case because the
separations between neighbouring energy levels increases at low
$\delta$-barrier ({\it i.e.} small $g$'s). At large $g$, this implies
a greater probability to excite higher-energy modes at small
temperature. Evidently, the violation of CHSH inequality becomes very
sensitive to temperature variations once the thermal energy is
comparable to the energy-level separation.
\begin{figure}[b]
\begin{center}
\includegraphics[width=7in]{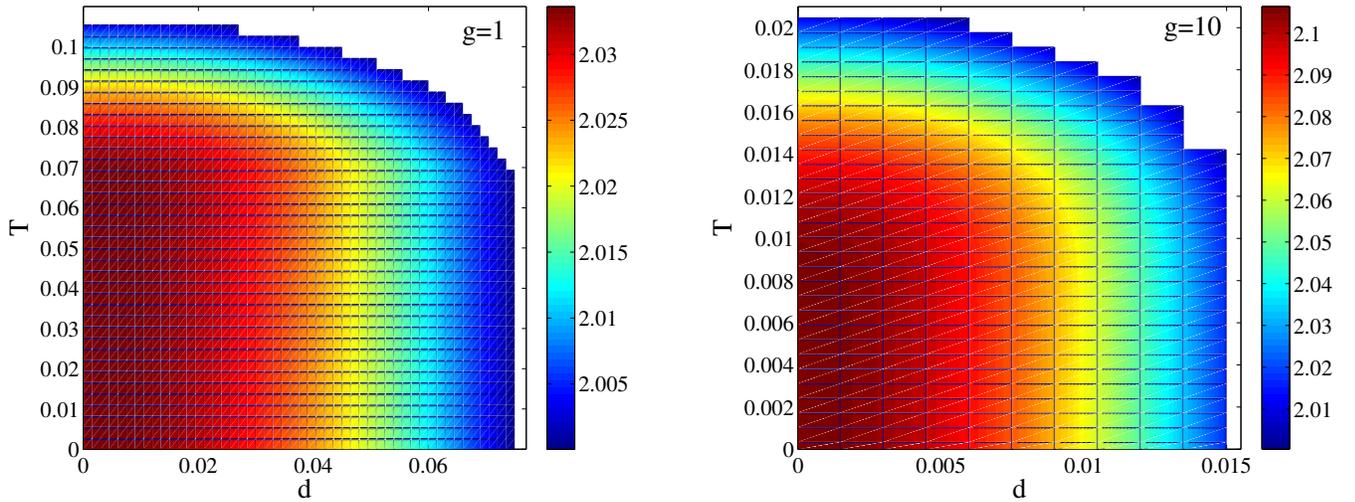}
\end{center}
\caption{Degree of CHSH violation against trap-separation and
  temperature (in units of $\hbar \omega / k_{B} $) for interaction strengths of $g=1$ and $g=10$. The
  change in temperature has a greater impact on the two particles at
  greater interaction energies (note the different scalings in the two plots).}
\label{fig:CHSHFT}
\end{figure}

We conclude this Section by sketching a strategy for the
reconstruction of the atomic Wigner function for a non-locality
test following the approach suggested by Lutterbach and
Davidovich~\cite{Lutterbach:97}. The
key is mapping the information encoded in the external degree of
freedom of one of the trapped atoms into specific internal state of
the atom itself, which can then be efficiently read out. For the sake
of argument, let us for the moment address the case of a single atom
and label the logical states of the qubit as
$\{|{\uparrow}\rangle,|{\downarrow}\rangle\}$. Physically, they could
be two quasi-degenerate metastable ground states of a three-level
$\Lambda$-like model and transitions from each ground state to
the excited level of the $\Lambda$ model will induce motional state-dependent
sidebands are induced on $|{\uparrow}\rangle$ and
$|{\downarrow}\rangle$. The transition between different motional
states of the atom can thus be induced by properly tuned stimulated
Raman passages connecting two different sidebands of the ground-state
doublet, as described in~\cite{Paternostro:04}, in a way so as to
mimic the dynamics intertwining motional degrees of freedom and
internal ones in trapped ions. Such processes can be performed with an
almost ideal efficiency. Working in an appropriate Lamb-Dicke
limit (where the recoil energy due to the {\it kicks} induced by the
coupling between atomic levels and light is much smaller than the
ground-state energy of the motional mode), it is possible to relate
the difference $P_\uparrow-P_\downarrow$ between the probability of
finding the atom in $|{\uparrow}\rangle$ or $|{\downarrow}\rangle$,
respectively, to the expectation value of the displaced parity
operator and thus, in turn, to the value of the Wigner function at a
given point of the phase space~\cite{Lutterbach:97}. Such a difference
in probability can be effectively measured by means of routinely
implemented high-efficiency fluorescence light-based detection
methods~\cite{Monroe:96}. In order to reconstruct the two-atom Wigner
function, it would be sufficient to collect signals from both the
atoms undergoing similar reconstruction protocols and appropriately
putting together the statistical data gathered. In doing this, distinguishability of the 
signals collected from the two particles can be attained by using fluorescence cycles of 
different frequencies, one per particle. In this way, one can distinguish the statistics associated with 
a specific particle without the need of separating the corresponding traps by a large distance.

%%%%%%%%%%%%%%%%%
\section{Effects of dissipation}
\label{sec:LossyDetector}

Let us finally discuss the influence of a general loss mechanisms, one
per atomic mode, that may affect the two-atom state due to finite-time
coherence of the external degrees of freedom. Such a lossy process can be effectively modelled 
considering each atomic vibrational mode as in contact with a background bath of bosons (due, for instance, to mode-mode coupling induced by an-harmonicity of the traps or position-to-electric-field coupling induced by stray electromagnetic fields in the proximity of the trapped particles)~\cite{bollinger}. The master equation arising from such a coupling can be then tackled by transforming it into a Fokker-Planck equation, which has exactly the same form as the one describing the propagation of an optical mode in a lossy channel~\cite{WallsMilburn}. Alternatively, our study
could equivalently be used so as to take into account the effects of a
finite-efficiency detection apparatus for the non-locality test
(although fluorescence-based methods have very high efficiency, boosted up with respect to single-photon detection efficiency by the large number of photons carried by the fluorescence signal, ideality of signal-collection is not yet achieved). Both
models can be abstractly yet rigorously described by considering a
simple {\it beam-splitter model} as follows: assuming low temperature
environments allows us to describe them as two independent zero-$T$
bosonic baths, each prepared in its collective vacuum state. We call
$A$ ($B$) the environmental bath affecting mode $1$ ($2$). The Wigner
function of the vacuum state of each is
\begin{equation}
  W_{0}(\mu_{k})=\frac{2}{\pi}e^{-2|\mu_{k}|^{2}}~~~(k=A,B),
\end{equation}
where $\mu_{k}=\frac{{x}_{k}+i{p}_{k}}{\sqrt{2}}$
are the quadrature variables of the environmental modes. The
interaction between the signal mode $j$ and its environment is
modelled as a mixing at a beam splitter having reflectivity
$\eta_k$. For simplicity and without affecting the generality of our
discussions, we assume the reflectivity to be equal in both modes,
$\eta_k=\eta$. In phase space, the state of the signal mode after the
interaction and after tracing over the environmental degrees of
freedom is described by the convolution
\begin{eqnarray}
\label{modified}
W^{\eta}(x_{1},x_{2},p_{1},p_{2})&{=}\!\int dx_{A} dx_{B} dp_{A} dp_{B}W(\tilde{x}_{1},\tilde{p}_{1},\tilde{x}_{2},\tilde{p}_{2})\nonumber\\
&\quad\times W_{0}(\tilde{x}_{A},\tilde{p}_{A})W_{0}(\tilde{x}_{B},\tilde{p}_{B}),
\end{eqnarray} 
where we have introduced the transformed variables
\begin{eqnarray}
\tilde{x}_{j}&=\sqrt{\eta}\,x_{j}-\sqrt{1-\eta}\,x_{k},~~\tilde{x}_{k}=\sqrt{\eta}\,x_{k}+\sqrt{1-\eta}\,x_{j},\\
\tilde{p}_{j}&=\sqrt{\eta}\,p_{j}-\sqrt{1-\eta}\,p_{k},~~\tilde{p}_{k}=\sqrt{\eta}\,p_{k}+\sqrt{1-\eta}\,p_{j}
\end{eqnarray}
and one should take $k=A$ ($B$) if $j=1$ ($2$).
Eq.~(\ref{modified}) is evaluated numerically and used to test violation of the CHSH inequality against $\eta$. Needless to say, the effect of losses (or detection inefficiencies) is to reduce the degree of violation of the CHSH inequality, as shown by the solid blue lines in Fig.~\ref{fig:CHSHFT}. The same trend highlighted before regarding resilience of non-locality properties for lower values of $g$ is retrieved here.
\begin{figure}
\begin{center}
\includegraphics[width=\linewidth]{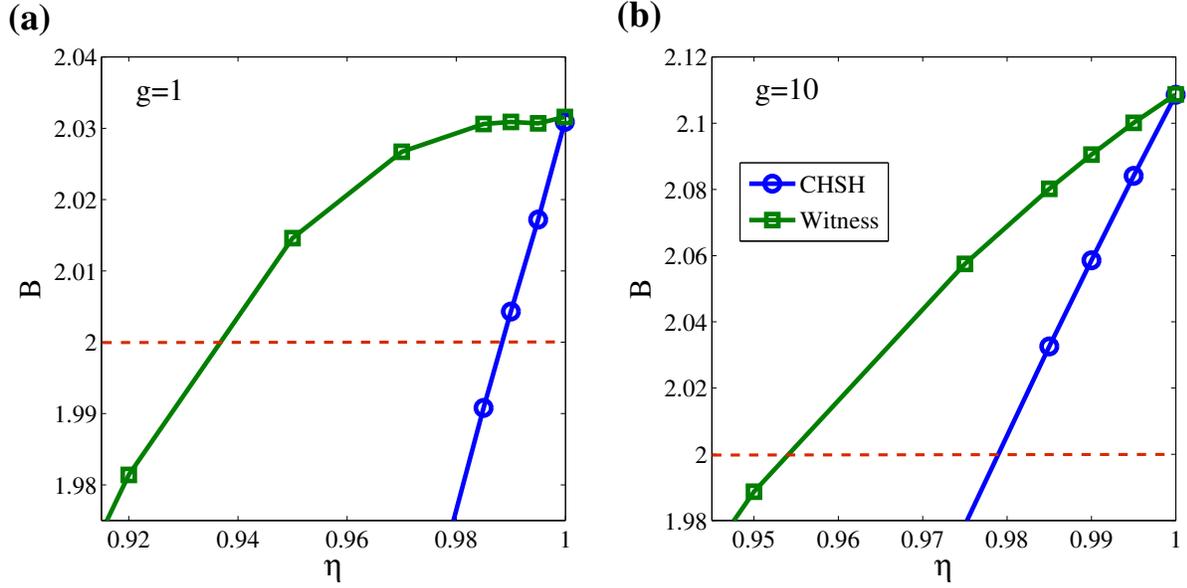}
\end{center}
\caption{Plots of CHSH violation and entanglement witness for
  interaction strengths of $g=1$ (panel {\bf (a)}) and $g=10$ (panel {\bf (b)})
  with inefficiency $\eta$. The violation of CHSH is seen to decay
  quickly as the detection becomes inefficient (blue lines), however
  the entanglement witness (green lines) shows the existence of
  entanglement for greater inefficiencies, with the lower value of $g$
  having more resilience to the losses.}
\label{fig:CHSHFT}
\end{figure}

It is therefore highly desirable to design viable strategies for a
more robust analysis of non-locality. A step in this direction has
been recently performed in Ref.~\cite{Lee:10} with the proposal of a
robust entanglement witness based on a CHSH-like inequality that shows
resilience with respect to losses/detection inefficiencies of the form
considered here. Following the derivation provided by Lee {\it et
  al.}~\cite{Lee:10}, one can see that for separable bipartite states
and loss rate/detection inefficiency $\eta$, the following
inequalities hold
\begin{eqnarray}
\mid\langle \mathcal{{W}}_{\eta >\frac{1}{2}}\rangle\mid &{=}| \frac{\pi^{2}}{4\eta^{2}}[W^{\eta}(0,0){+}W^{\eta}(0,-\sqrt{\mathcal{J}}){+}W^{\eta}(\sqrt{\mathcal{J}},0){-}W^{\eta}(\sqrt{\mathcal{J}},-\sqrt{\mathcal{J}}) ]\nonumber \\
&+\frac{\pi(\eta-1)}{\eta^{2}}[W^{\eta}_{\alpha}(0)+W^{\eta}_{\beta}(0)]+2(1-\frac{1}{\eta})^{2} | \leq 2, \nonumber \\
\mid\langle \mathcal{{W}}_{\eta \leq \frac{1}{2}}\rangle\mid &{=} | \pi^{2}[W^{\eta}(0,0)+W^{\eta}(0,-\sqrt{\mathcal{J}})+W^{\eta}(\sqrt{\mathcal{J}},0)-W^{\eta}(\sqrt{\mathcal{J}},-\sqrt{\mathcal{J}}) ]\nonumber \\
&-2\pi[W^{\eta}_{1}(0)+W^{\eta}_{2}(0)]+2 | \leq 2. 
\end{eqnarray}
Here, $W^{\eta}(a,b)$ is the two-mode Wigner function calculated in
Eq.~(\ref{modified}) and $W^{\eta}_{1,2}$ are its single-mode
marginals. For the case of perfect detectors ($\eta=1$) the inequality
becomes equivalent to (\ref{eqn:chsh}). It is apparent that any
violation of this inequality for $\eta<1$ ensures the violation of the
CHSH-inequality in the presence of the unitary case as well, thus such
witness can be used effectively for detecting entanglement in the
presence of noise. From the results shown in Fig.~\ref{fig:CHSHFT} one can see
that, while the CHSH inequality violation is lost for $\eta=0.98$ at
$g=10$, the entanglement witness still violates it at $\eta\simeq0.95$,
which is a small yet significant improvement. It is important to
notice that current avalanche photodiodes used to collect fluorescence
have quantum efficiencies exactly in this range. Interestingly the
entanglement witness for $g=1$ is violated for smaller $\eta$, echoing
the trend noticed for the CHSH at zero and non-zero temperature: lower
interaction strengths give rise to states more resilient to influences
from the environment.

\section{Conclusions}
\label{sec:Conclusions}

We have investigated CV entanglement in a system of two interacting
bosons in separate harmonic trapping potentials under a variety of
conditions. We have found that the von Neumann entropy shows strong correlations at zero
temperature and shown violation of local realistic theories for a wide range
of relevant parameters. An interesting and rather counterintuitive
behavior has been observed, even at non-zero temperature, for the
whole range of interaction strengths analysed. We have related the multiple facets of both the revealed
non-locality and the von Neumann entropy to the details of
the coupling model used in this work and the corresponding spectrum of the system. 

Finally, we have included the effects of general non-idealities (such as dissipative losses
affecting the motional degrees of freedom of the trapped atoms or
inefficient detectors), demonstrating the fragility of the
atomic non-locality. In order to circumvent such a hindrance, we have
shown that some improvements can come from the use of a recently
proposed entanglement witness that fits very well with the general
approach put forward here. We hope that the realistic nature of our
proposal functions as a significant
model to test non-classicality of massive systems. Since it is 
rather close to state of the art experimental possibilities we expect it to spur experimental
interest in the study of non-classical behaviour of simple
low-dimensional atomic systems under non-ideal working conditions.

\section{Acknowledgements}
We acknowledge financial support from the Irish Research Council for
Science and Engineering under the Embark initiative RS/2009/1082 and
Science Foundation Ireland under grants no. 05/IN/I852 and 05/IN/I852
NS. JG would like to thank the National Research Foundation and
Ministry of Education of Singapore for support and Mr.~G.~Vacanti for
interesting discussions. MP is grateful to Dr.~Seungwoo Lee and
Prof.~Hyunseok Jeong for invaluable comments and thanks the UK EPSRC
for financial support (EP/G004579/1).

\end{document}